\documentclass[10pt, aps, prd,
reprint, 
notitlepage, nofootinbib, longbibliography, showpacs, letter,
superscriptaddress,
]{revtex4-1}

\usepackage{pstricks,pst-fractal}
\usepackage{amsmath,amssymb,bm}
\usepackage{graphicx}
\usepackage{verbatim}
\usepackage[breaklinks=true,colorlinks=true,urlcolor=blue,linkcolor=red,citecolor=red]{hyperref}

\begin{document}

\title{Casimir energy of Sierpinski triangles}

\author{K. V. Shajesh}
\email{kvshajesh@gmail.com} \homepage{http://www.physics.siu.edu/~shajesh}
\affiliation{Department of Physics, Southern Illinois University--Carbondale,
Carbondale, Illinois 62901, USA}
\affiliation{Department of Energy and Process Engineering,
Norwegian University of Science and Technology, N-7491 Trondheim, Norway}

\author{Prachi Parashar}
\email{prachi.parashar@ntnu.no}
\homepage{https://www.ntnu.edu/employees/prachi.parashar}
\affiliation{Department of Energy and Process Engineering,
Norwegian University of Science and Technology, N-7491 Trondheim, Norway}

\author{In\'{e}s Cavero-Pel\'{a}ez}
\email{cavero@unizar.es} \homepage{http://cud.unizar.es/cavero}
\affiliation{Centro Universitario de la Defensa (CUD),
Zaragoza 50090, Spain}

\author{Jerzy Kocik}
\email{jkocik@siu.edu} \homepage{http://lagrange.math.siu.edu/Kocik/jkocik.htm}
\affiliation{Department of Mathematics,
Southern Illinois University--Carbondale, Carbondale, Illinois 62901, USA}

\author{Iver Brevik}
\email{iver.h.brevik@ntnu.no}
\homepage{http://folk.ntnu.no/iverhb}
\affiliation{Department of Energy and Process Engineering,
Norwegian University of Science and Technology, N-7491 Trondheim, Norway}

\date{\today}

\begin{abstract}
Using scaling arguments and the property of self-similarity we derive 
the Casimir energies of Sierpinski triangles and Sierpinski rectangles.
The Hausdorff-Besicovitch dimension (fractal dimension) of the Casimir 
energy is introduced and the Berry-Weyl conjecture is discussed for 
these geometries. We propose that for a class of fractals, comprising
of compartmentalized cavities, it is possible to establish a finite
value to the Casimir energy even while the Casimir energy of the
individual cavities consists of divergent terms.
\end{abstract}

\maketitle
\section{Introduction}

Weyl's law~\cite{weyl}, which was originally discussed for 
the spectral distribution of
the modes allowed inside a Dirichlet cavity, when extended for the 
Casimir energy per unit length ${\cal E}(a)$ of a polygonal
cylindrical cavity with a single characteristic scale $a$,
in natural units of $\hbar=c=1$, states that
\begin{equation}
{\cal E}(a) = \frac{b_c}{a^2} + \lim_{\tau\to 0} \frac{1}{\tau^2}
\left[ b_2 A\left( \frac{a}{\tau} \right) + b_1 P\left( \frac{a}{\tau} \right)
+ b_0 C\left( \frac{a}{\tau} \right) \right], \hspace{5mm}
\label{weyl-cas}
\end{equation}
where the coefficients of the divergent terms, $A(x)$, $P(x)$, and $C(x)$,
scale like the area of the cavity, the perimeter of the cavity,
and the corner angles of the cavity, respectively. That is,
\begin{equation}
A(x) \propto x^2, \quad P(x) \propto x^1, \quad C(x) \propto x^0. 
\end{equation} 
Parameters $b_c$, $b_2$, $b_1$, and $b_0$ in Eq.\,(\ref{weyl-cas}) 
are dimensionless constants. The parameter $\tau$ 
is a temporal point-splitting cutoff parameter 
introduced in the calculation to regulate the divergences.

Berry conjectured~\cite{Berry1979ps}, 
again in the context of spectral distribution,
that for fractal cavities the Weyl law maintains
the form of Eq.\,(\ref{weyl-cas}) with the only difference that the
coefficients of the divergent terms, $A(x)$, $P(x)$, and $C(x)$,
scale like the Hausdorff-Besicovitch dimension (fractal dimension) of 
the area of the cavity, the perimeter of the cavity,
and the corner angles of the cavity, respectively. That is,
\begin{equation}
A(x) \propto x^{\delta_2}, \quad P(x) \propto x^{\delta_1}, 
\quad C(x) \propto x^{\delta_0},
\label{def-dpp}
\end{equation}
where $\delta_2$ is the fractal dimension of the area of the cavity,
$\delta_1$ is the fractal dimension of the perimeter of the cavity,
and $\delta_0$ is the fractal dimension of the corner angles of the cavity.

\begin{figure}[t]
\includegraphics{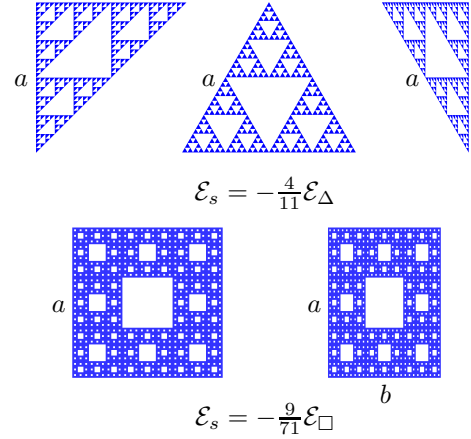}
\caption{Gallery of Sierpinski cylinders with Casimir energies
per unit length ${\cal E}_s$
for the five integrable cylinders studied in Ref.\,\cite{Abalo:2010ah}.
Top row: An isosceles right triangle with the equal sides of length $a$,
an equilateral triangle of side length $a$, and
a hemiequilateral triangle with length of hypotenuse $a$.
Bottom row: A square of side length $a$, and
a rectangle of side lengths $a$ and $b$.
The Casimir energy per unit length of a Sierpinski triangle is $-4/11$
times the Casimir energy per unit length of the respective triangle,
${\cal E}_\Delta$,
and the Casimir energy per unit length of a Sierpinski rectangle or
square is $-9/71$ times the Casimir energy per unit length
of the respective rectangle or square, ${\cal E}_\Box$. 
The Casimir energy per unit length, ${\cal E}_\Delta$ and ${\cal E}_\Box$,
for the five integrable cylinders are summarized in Table~\ref{tab-ectr}.
}
\label{fig-sierpinski-gallery}
\end{figure}%

It is, then, not a long shot to envision that
the Casimir energy per unit length of a fractal cavity
need not scale like the inverse square of length. Thus, presuming that 
the energy scales like $a^{\delta_c}$, we can generalize
Weyl's law in Eq.\,(\ref{weyl-cas}) as
\begin{equation}
{\cal E}(a) = b_c\,a^{\delta_c} + \lim_{\tau\to 0} \tau^{\delta_c}
\left[ b_2 A\left( \frac{a}{\tau} \right) + b_1 P\left( \frac{a}{\tau} \right)
+ b_0 C\left( \frac{a}{\tau} \right) \right], \hspace{5mm}
\label{gen-BWc}
\end{equation}
where $\delta_c$ is the fractal dimension of the Casimir energy per unit length
of the fractal cavity.

The central theme of this paper is to use the scaling arguments
and the property of self-similarity introduced 
in Ref.~\cite{Shajesh:2016yfq} to derive the Casimir energies
of Sierpinski triangles and Sierpinski rectangles.
We also introduce a class of fractals for which the energy
does not scale as inverse length square, which leads us
to introduce a fractal dimension for the Casimir energy.
One usually associates fractal dimensions to geometrical quantities
like perimeter and area, but being able to introduce
fractal dimensions to Casimir energy for a class of fractals
directly relates to the conventional
wisdom that Casimir energy of cavities, satisfying perfectly
conducting boundary conditions (or Dirichlet boundary conditions
for scalar fields), is purely geometrical.
It should not be very surprising, because energy has been shown to
exhibit fractal nature before.
For example, the Hofstadter butterfly is a fractal that represents
the energy of Bloch electrons in a magnetic field~\cite{Hofstadter:1976fmi}.

\begin{table}
\begin{tabular}{|l|c|c|c|}
\hline
Cross section & Dirichlet & Neumann & EM \\
\hline
Equilateral Tr.
\rule{0pt}{5ex}
& $\dfrac{0.0237}{a^2}$ & $-\dfrac{0.0613}{a^2}$ & $-\dfrac{0.0375}{a^2}$
\\[3mm]
Hemiequilateral Tr.
& $\dfrac{0.0756}{a^2}$ & $-\dfrac{0.0944}{a^2}$ & $-\dfrac{0.0187}{a^2}$
\\[3mm]
Isosceles Tr.
& $\dfrac{0.0263}{a^2}$ & $-\dfrac{0.0454}{a^2}$ & $-\dfrac{0.0190}{a^2}$
\\[3mm]
Square
& $\dfrac{0.00483}{a^2}$ & $-\dfrac{0.0429}{a^2}$ & $-\dfrac{0.0381}{a^2}$
\\[3mm]
\hline
Rectangle \rule{0pt}{4ex}
& \multicolumn{3}{|c|}{Refer Ref.~\cite{Abalo:2010ah}.} \\[2mm]
\hline
\end{tabular}
\caption{
Casimir energy per unit length for cylinders of five cross sections
from Ref.~\cite{Abalo:2010ah}, referred to as
${\cal E}_\Delta$ and ${\cal E}_\Box$ in this paper.
The cutoff independent finite part is presented.
The numbers correspond to the constant $b_c$ in
Eq.\,(\ref{weyl-cas}) for the respective cross sections,
presented here to three significant digits without rounding.
The second, third, and fourth,
columns correspond to the boundary conditions imposed on the fields.}
\label{tab-ectr}
\end{table}

We shall consider an equilateral triangle even though
most of our discussion holds true for an arbitrary triangle.
The Casimir energy of an equilateral triangular cylinder
on which a scalar field satisfies Dirichlet boundary conditions
was calculated exactly, in closed form, in Ref.~\cite{Abalo:2010ah}.
This involves the Casimir energy of five cylindrical cross sections, namely
an equilateral triangle, a hemiequilateral triangle,
an isosceles right triangle, a square, and a rectangle,
see Fig.~\ref{fig-sierpinski-gallery}. For all five geometries
the authors of Ref.~\cite{Abalo:2010ah} have shown that the Casimir energy 
per unit length obeys the Weyl law in Eq.\,(\ref{weyl-cas}).
Using Ref.~\cite{Abalo:2010ah}, the Casimir energy per unit length of an 
equilateral triangle is described by the parameters
\begin{equation}
b_c = -\frac{1}{72} \left[ \frac{\sqrt{3}}{9} \left\{
\psi^{(1)}\left({\textstyle \frac{1}{3} }\right) 
-\psi^{(1)}\left({\textstyle \frac{2}{3} }\right)
\right\} -\frac{8}{\pi} \zeta(3) \right],
\end{equation} 
with a numerical value $b_c\sim 0.0237188$,
\begin{equation}
b_2=\frac{3\sqrt{3}}{8\pi^2}, \quad b_1=-\frac{3}{8\pi}, 
\quad b_0=\frac{1}{6\pi},
\end{equation}
of Eq.\,(\ref{weyl-cas}), 
where $\psi^{(m)}(z)$ is the polygamma function of order $m$
and $\zeta(z)$ is the Riemann zeta function.
The above evaluation was achieved using the mode summation method,
which presumes that the Casimir energy of a closed Dirichlet cavity 
is completely determined by the modes in the interior of the cavity 
alone~\cite{Abalo:2010ah}.
This should be contrasted with field theoretic methods of 
Lifshitz et al.~\cite{Dzyaloshinskii:1961fwd} 
and Schwinger et al.~\cite{Schwinger:1977pa}
that incorporate both the interior and exterior modes in the evaluation.
The Casimir energies for the five geometries of Ref.~\cite{Abalo:2010ah}, 
for empty triangles and rectangles or squares, due to interior modes only,
referred to as ${\cal E}_\Delta$ and ${\cal E}_\Box$ in this paper,
are summarized in Table~\ref{tab-ectr}.

Before we proceed with our discussion, we present the results 
for the finite part of the Casimir energy 
for each of the five geometries presented in 
Fig.~\ref{fig-sierpinski-gallery}, which are expressed in terms of
${\cal E}_\Delta$ and ${\cal E}_\Box$ of Table~\ref{tab-ectr}.
The expression ${\cal E}_s =-4{\cal E}_\Delta/11$ is universal for
all Sierpinski triangles, and
${\cal E}_s =-9{\cal E}_\Box/71$ is universal for all Sierpinski rectangles,
and are not restricted to the five geometries of Ref.~\cite{Abalo:2010ah}.
However, we often confine the analysis to the five geometries of 
Ref.~\cite{Abalo:2010ah}, because exact expressions were derived for
the Casimir energy per unit length for these geometries there.

\section{Sierpinski triangle}

\begin{figure}[ht]
\includegraphics{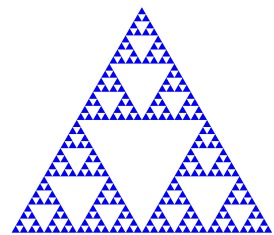}
\caption{Sierpinski triangle. The white regions in the interior are 
triangular cavities, each of which contributes to the Casimir energy
per unit length of the Sierpinski triangle. The matter bounding each of the 
triangles (in blue) are perfectly conducting for the case of
electromagnetic fields.}
\label{fig-sierT}
\end{figure}

The Sierpinski triangle is self-similar. 
That is, it consists of copies of the scaled-down versions of itself.
Figure~\ref{fig-sierT} shows the Sierpinski triangle of side length $a$,
which may be viewed as comprised of three Sierpinski triangles 
of side length $a/2$.

\subsection{Area and $\delta_2$}

Using the self-similarity of a Sierpinski triangle we can write the
following recursion relation for the area $A_s(a)$ of the cavities inside 
a Sierpinski triangle:
\begin{equation}
A_s(a) = A_\Delta\left(\frac{a}{2}\right) +3A_s\left(\frac{a}{2}\right),
\label{area-st-re}
\end{equation}
where $A_\Delta(a)$ is the area of an equilateral triangle of side length $a$.
Using Eq.\,(\ref{area-st-re}) recursively in itself we obtain the series
\begin{equation}
A_s(a) = A_\Delta\left(\frac{a}{2}\right) +3\,A_\Delta\left(\frac{a}{2^2}\right)
+3^2\,A_\Delta\left(\frac{a}{2^3}\right) +\ldots.
\end{equation}
Then, using the scaling relation of the area of a triangle,
\begin{equation}
A_\Delta\left(\frac{a}{2}\right) = \frac{1}{2^2} A_\Delta(a),
\end{equation}
we obtain the area of the Sierpinski triangle in Fig.~\ref{fig-sierT}
to be exactly equal to the area of a triangle. That is,
\begin{equation}
A_s(a) = A_\Delta(a).
\label{ast=att}
\end{equation}
Since the area of a triangle $A_\Delta(a)$ scales like $a^2$,
dimensional analysis of Eq.\,(\ref{ast=att}) implies that the
fractal dimension of the area, defined in Eq.\,(\ref{def-dpp}),
of the Sierpinski triangle is
\begin{equation}
\delta_2=2.
\end{equation}

\subsection{Perimeter and $\delta_1$}

Similarly, the interior perimeter of the cavities inside a Sierpinski triangle
satisfies the recursion relation
\begin{equation}
P_s(a) = P_\Delta\left(\frac{a}{2}\right) +3P_s\left(\frac{a}{2}\right),
\label{peri-st-re}
\end{equation}
where $P_\Delta(a)$ is the interior perimeter of a triangular cavity 
of side length $a$ and $P_s(a)$ is the sum of the interior perimeter of
all the individual cavities constituting the Sierpinski triangle.
The series constructed from the recursion relation, after using
the scaling argument for $P_\Delta(a)$, is divergent and leads to
\begin{equation}
P_s(a) = -P_\Delta(a)
\label{peri-tri-st}
\end{equation}
after using the divergent sum
\begin{equation}
\frac{1}{2} +\frac{3}{2^2} +\frac{3^2}{2^3} +\ldots = -1,
\end{equation}
which can be deduced using the property of self-similarity of the
series~\cite{Hardy-b1109376}.
Ignoring the counterintuitive nature of a negative perimeter,
we learn from Eq.\,(\ref{peri-tri-st}) that the
fractal dimension for the perimeter, defined in Eq.\,(\ref{def-dpp}),
of the Sierpinski triangle in Fig.~\ref{fig-sierT} is
\begin{equation}
\delta_1=1,
\end{equation}
because the perimeter of a triangle $P_\Delta(a)$ scales like $a$.

\subsection{Corner angle and $\delta_0$}

The interior corner angles of a Sierpinski triangle satisfy
\begin{equation}
C_s(a) = C_\Delta\left(\frac{a}{2}\right) +3C_s\left(\frac{a}{2}\right),
\label{cor-st-re}
\end{equation}
which leads to the series
\begin{equation}
C_s(a) = C_\Delta\left(\frac{a}{2}\right) +3\,C_\Delta\left(\frac{a}{2^2}\right)
+3^2\,C_\Delta\left(\frac{a}{2^3}\right) +\ldots.
\end{equation}
The corner term for a triangle is given by~\cite{Abalo:2010ah}
\begin{equation}
C_\Delta (a) = \sum_i \left( \frac{\pi}{\alpha_i} -\frac{\alpha_i}{\pi} \right),
\end{equation}
where $\alpha_i$ are the angles of a triangle, is independent of the scale $a$.
Thus, we can derive
\begin{equation}
C_s(a) = -\frac{1}{2} C_\Delta(a),
\label{cor-tri-st}
\end{equation}
after using the divergent sum $1+3+3^2+\ldots =-1/2$.
We learn from Eq.\,(\ref{cor-tri-st}) that the
fractal dimension for the corner angle of the Sierpinski triangle
in Fig.~\ref{fig-sierT} is
\begin{equation}
\delta_0=0,
\end{equation}
because the corner angle of a triangle $C_\Delta(a)$ is scale independent.

\subsection{Casimir energy and $\delta_c$}

Using the decomposition of Casimir energies into single-body energy
and the respective interaction energy between the bodies~\cite{Shajesh:2011ef},
the Casimir energy per unit length of a Sierpinski triangle ${\cal E}_s(a)$
can be decomposed as
\begin{equation}
{\cal E}_s(a) = {\cal E}_\text{int}\left(\frac{a}{2}\right) 
+ 3{\cal E}_s\left(\frac{a}{2}\right),
\label{st-ed3}
\end{equation}
where $3{\cal E}_s(a/2)$ is the single-body Casimir energy of three
Sierpinski triangles of side $a/2$ in Fig.~\ref{fig-sierT}, and
${\cal E}_\text{int}(a)$ is the interaction energy between the 
three Sierpinski triangles.
Arguably, in general, the interaction energy ${\cal E}_\text{int}$
depends on both the interior and exterior modes. But, the Casimir energies
of the cavities we are considering are all due to interior modes.
Thus, for consistency, we shall presume the interaction energy 
involves only the interior modes. We shall further justify the 
consistency of this presumption in the following discussion.

Using Eq.\,(\ref{st-ed3}) recursively we obtain the series
\begin{equation}
{\cal E}_s(a) = {\cal E}_\text{int}\left(\frac{a}{2}\right) 
+ 3\,{\cal E}_\text{int}\left(\frac{a}{2^2}\right) 
+ 3^2\,{\cal E}_\text{int}\left(\frac{a}{2^3}\right) +\ldots.
\label{steint}
\end{equation}
Thus, the evaluation of the Casimir energy reduces to computing the 
interaction energy ${\cal E}_\text{int}$. 

\begin{figure}[ht]
\includegraphics{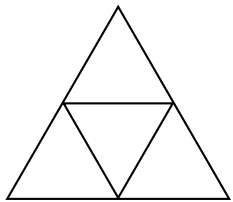}
\caption{Four triangles constituting the Sierpinski triangle.}
\label{fig-fourT}
\end{figure}%

Dirichlet boundary conditions requires a scalar field to be zero
on the boundary. This restriction essentially separates the 
physical phenomena on the two sides of the boundary.
Thus, the modes and the associated physical phenomena inside 
Dirichlet cavities are essentially independent of its surroundings. 
Extending this argument to Sierpinski triangles we learn that 
the interaction energy between two or more Sierpinski triangles is 
independent of the internal structure of each of them.
We can thus infer that the interaction energy of the three 
Dirichlet Sierpinski triangles in Fig.~\ref{fig-sierT}
is identical to the interaction energy of the three Dirichlet triangles
in Fig.~\ref{fig-fourT}.
We can determine the total energy of the four triangles in 
Fig.~\ref{fig-fourT} in two independent methods.
In the first method we argue that the energy is the sum of the four
triangular cavities, $4{\cal E}_\Delta\left(\frac{a}{2}\right)$.
In the second method we argue that the total energy is the sum 
of the energies of the three outer triangles,
$3{\cal E}_\Delta\left(\frac{a}{2}\right)$,
plus the interaction energy ${\cal E}_\text{int}(a/2)$ 
between the three triangles. That is,
\begin{equation}
4{\cal E}_\Delta\left(\frac{a}{2}\right) 
=3{\cal E}_\Delta\left(\frac{a}{2}\right) 
+ {\cal E}_\text{int}\left(\frac{a}{2}\right). 
\end{equation}
This immediately suggests that the interaction energy of three outer
triangles is completely given by the energy of the inner triangle, 
\begin{equation}
{\cal E}_\text{int}\left(\frac{a}{2}\right) 
= {\cal E}_\Delta\left(\frac{a}{2}\right) = 2^2 {\cal E}_\Delta(a),
\label{est-eint4}
\end{equation}
where in the second equality we used the
fact that the Casimir energy of a triangle ${\cal E}_\Delta(a)$ scales
like the inverse square of length.

Using Eq.\,(\ref{est-eint4}) in Eq.\,(\ref{steint}) we derive the Casimir
energy of the Dirichlet Sierpinski triangle in terms of the Casimir energy
of the equilateral triangle as 
\begin{equation}
{\cal E}_s(a) = -\frac{4}{11} {\cal E}_\Delta(a)
\label{cEs=cEc}
\end{equation}
using the divergent sum
\begin{equation}
1 +12 +12^2 +\ldots = -\frac{1}{11}.
\end{equation}
Since the energy per unit length of a triangle ${\cal E}_\Delta(a)$
scales like inverse length square, Eq.\,(\ref{cEs=cEc}) implies that
the energy per unit length of the Sierpinski triangle also scales
similarly, that is,
\begin{equation}
\delta_c=-2.
\end{equation}

The Casimir energy per unit length for the Sierpinski extension of 
a cylinder with arbitrary triangular cross section will also be 
given using Eq.\,(\ref{cEs=cEc}).
The expression for ${\cal E}_\Delta$, without the cutoff dependent part,
for the three cylinders with triangular cross sections, for which closed-form
solutions has been achieved, has been summarized in Table~\ref{tab-ectr}.

\section{Inverse Sierpinski triangle}

\begin{figure}[ht]
\includegraphics{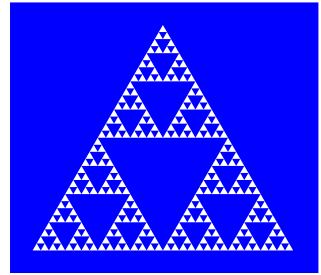}
\caption{Inverse Sierpinski triangle.
It is obtained from the Sierpinski triangle in Fig.~\ref{fig-sierT} 
by swapping the empty space with perfectly conducting material there.
The blue region extends to infinity.}
\label{fig-inverse-sierT}
\end{figure}

We define the inverse Sierpinski triangle as the object obtained by 
swapping the empty space with the perfectly conducting material
in the Sierpinski triangle.
In Fig.~\ref{fig-sierT} this is obtained by swapping the
white color representing empty space with blue color
representing perfectly conducting material.
See Fig.~\ref{fig-inverse-sierT}.
The outer region in Fig.~\ref{fig-inverse-sierT} is unbounded.

The area of the inverse Sierpinski triangle satisfies the relation
\begin{equation}
A_s(a) = 3^n\,A_s\left(\frac{a}{2^n}\right),
\label{area-ist}
\end{equation}
for any non-negative integer $n$.
Presuming that this area scales like $a^{\delta_2}$ we obtain the relation
\begin{equation}
A_s(a) = \frac{3^n}{(2^n)^{\delta_2}} A_s(a),
\label{frdim-ar}
\end{equation}
for any positive integer $n$.
The central idea of non-trivial fractal dimensions stems from 
Eq.\,(\ref{frdim-ar}) and its solutions. 
The trivial solution is $A_s(a)=0$,
which can be envisioned to be a possible scenario by extending 
Fig.~\ref{fig-inverse-sierT} (drawn to 5 iterations) to infinite iterations.
This trivial solution agrees with the notion that the perfectly conducting
material in Fig.~\ref{fig-inverse-sierT} fills all space in this limit.
But, Eq.\,(\ref{frdim-ar}) also admits a non-trivial solution, namely
\begin{equation}
1 = \left( \frac{3}{2^{\delta_2}} \right)^n, 
\end{equation}
for any positive integer $n$, which immediately implies that
\begin{equation}
\delta_2= \frac{\ln 3}{\ln 2} \sim 1.58496.
\end{equation}
The non-trivial solution here is probably a consequence of the non-trivial
solution of a divergent series as a regularized sum,
which has the trivial solution to be infinity~\cite{Hardy-b1109376}.

The perimeter also satisfies the relation in Eq.\,(\ref{area-ist})
with the areas $A$ now replaced with perimeters $P$. The corner angles also
satisfy Eq.\,(\ref{area-ist}).
Further, the Casimir energy per unit length also satisfies
Eq.\,(\ref{area-ist}). Thus, we learn that
\begin{equation}
\delta_c=\delta_2=\delta_1=\delta_0 =\frac{\ln 3}{\ln 2},
\end{equation}
unless we confine to the trivial solution that area, perimeter,
angles, and the energy per unit length are all zero.

The fact that the fractal dimensions of all the relevant
physical quantities, the area, the perimeter,
the corner angle, and the Casimir energy per unit length, scale the same way,
in conjunction with the generalized Berry conjecture of Eq.\,(\ref{gen-BWc})
implies that there are no divergent terms in the Casimir energy per unit length
of the inverse Sierpinski triangle. That is,
\begin{eqnarray}
{\cal E}(a) &=& b_c\,a^{\delta_c} + \lim_{\tau\to 0} \tau^{\delta_c}
\left[ b_2^\prime \left( \frac{a}{\tau} \right)^{\delta_c}
+ b_1^\prime \left( \frac{a}{\tau} \right)^{\delta_c}
+ b_0^\prime \left( \frac{a}{\tau} \right)^{\delta_c} \right], 
\nonumber \\
&=& (b_c +b_2^\prime +b_1^\prime +b_0^\prime) a^{\delta_c},
\label{no-div-C}
\end{eqnarray}
where $b_2^\prime$, $b_1^\prime$, and $b_0^\prime$ are redefined constants
with respect to the constants $b_2$, $b_1$, and $b_0$,
in Eq.\,(\ref{gen-BWc}), to accommodate numerical constants inside 
$A(x)$, $P(x)$, and $C(x)$, respectively.

\section{Sierpinski carpet}

\begin{figure}[t]
\includegraphics[angle=45]{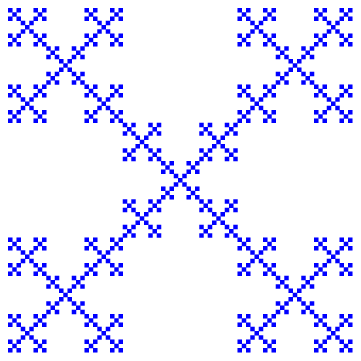}
\caption{Vicsek fractal.}
\label{fig-vicsekF}
\end{figure}%

The Sierpinski carpet, or a Sierpinski rectangle or square,
is the rectangular version of a Sierpinski triangle,
see the rectangle and square version in Fig.~\ref{fig-sierpinski-gallery}.
A Sierpinski carpet satisfies the energy decomposition
\begin{equation}
{\cal E}_c(a) = {\cal E}_\text{int}\left(\frac{a}{3}\right)
+ 8{\cal E}_c\left(\frac{a}{3}\right),
\end{equation}
which leads to ${\cal E}_c(a) = -{\cal E}_\text{int}(a/3)/71$. 
Again, to be consistent with the fact that we are including only
the interior modes, the interaction energy ${\cal E}_\text{int}(a/3)$ 
is equal to the Casimir energy of a square enclosed between the eight 
surrounding squares. Thus, we have
\begin{equation}
{\cal E}_\text{int}\left(\frac{a}{3}\right)
={\cal E}_\Box \left(\frac{a}{3}\right)
=3^2{\cal E}_\Box (a),
\end{equation}
where ${\cal E}_\Box (a)$ is the Casimir energy per unit length of
a square of side length $a$. Hence, the Casimir energy of the Sierpinski square
is evaluated as
\begin{equation}
{\cal E}_c(a) = -\frac{9}{71} {\cal E}_\Box (a).
\label{cEs=cEcRe}
\end{equation}
The calculation for the Sierpinski rectangle goes through the same procedure,
because the length and width have a fixed aspect ratio.
The Casimir energy per unit length for the Sierpinski extension of
a cylinder with arbitrary rectangular cross section will also be
given using Eq.\,(\ref{cEs=cEcRe}).
The expression for ${\cal E}_\Box$, without the cutoff dependent part,
for the cylinders with rectangular cross sections, for which closed-form
solutions have been found, are summarized in Table~\ref{tab-ectr}.

The vacuum energy of the inverse Sierpinski square or rectangle
satisfies the relation
\begin{equation}
{\cal E}_c(a) = 8^n{\cal E}_c\left(\frac{a}{3^n}\right),
\label{eiC=Re}
\end{equation}
which implies $\delta_c=\ln 8/\ln 3$. Since the area, the perimeter,
and the corner angles, satisfy the same relation,
of Eq.\,(\ref{eiC=Re}), we also learn that
$\delta_c =\delta_2 =\delta_1 =\delta_0$.
Thus, using the same arguments used to derive Eq.\,(\ref{no-div-C}),
the Casimir energy per unit length of the inverse Sierpinski
rectangle or square will not have divergent terms.

We can also extend our discussion to non-Sierpinski fractals.
The Vicsek fractal, illustrated in Fig.~\ref{fig-vicsekF}, is obtained by
starting from a square, dividing it into nine equal squares of one-third side,
and removing four of them. The inverse Vicsek fractal is obtained by
swapping the empty space with perfectly conducting material, such that 
the Casimir energy is given in terms of the energies of the individual
cavities. The vacuum energy of the inverse Vicsek fractal satisfies
the relation
\begin{equation}
{\cal E}_v(a) = 5^n{\cal E}_v\left(\frac{a}{3^n}\right),
\label{eiV=Re}
\end{equation}
and the area, the perimeter, and the corner angles, of the Vicsek fractal also
satisfy the same relation. Like in the case of the inverse Sierpinski triangle,
Eq.\,(\ref{eiV=Re}) for energy and the corresponding relations
admits the non-trivial solution
\begin{equation}
\delta_c =\delta_2 =\delta_1 =\delta_0=\frac{\ln 5}{\ln 3} \sim 1.46497.
\end{equation}
Thus, using the generalized Berry conjecture of Eq.\,(\ref{gen-BWc})
we can conclude that the Casimir energy per unit length of the inverse
Vicsek fractal will not have divergent terms.

\begin{table}
\begin{tabular}{|l|c|c|c|c|}
\hline
\hspace{10mm} Geometry & $\delta_2$ & $\delta_1$ & $\delta_0$ & $\delta_c$ \\
\hline
Sierpinski triangle & 2 & 1& 0& -2 \\[2mm]
Inverse Sierpinski triangle & $\dfrac{\ln 3}{\ln 2}$ 
& $\dfrac{\ln 3}{\ln 2}$ & $\dfrac{\ln 3}{\ln 2}$ 
& $\dfrac{\ln 3}{\ln 2}$ \\[4mm]
Sierpinski carpet & 2 & 1& 0& -2 \\[2mm]
Inverse Sierpinski carpet & $\dfrac{\ln 8}{\ln 3}$ 
& $\dfrac{\ln 8}{\ln 3}$ & $\dfrac{\ln 8}{\ln 3}$ 
& $\dfrac{\ln 8}{\ln 3}$ \\[4mm]
Inverse Vicsek fractal & $\dfrac{\ln 5}{\ln 3}$
& $\dfrac{\ln 5}{\ln 3}$ & $\dfrac{\ln 5}{\ln 3}$
& $\dfrac{\ln 5}{\ln 3}$ \\[4mm]
Koch snowflake & 2 & $\dfrac{\ln 4}{\ln 3}$ & 0 & ? \\[2mm]
\hline
\end{tabular}
\caption{Fractal dimensions of the area of cavities, $\delta_2$,
of the perimeter of cavities, $\delta_1$, of the corner angles of cavities,
$\delta_0$, and of the Casimir energy per unit length, $\delta_c$,
for a few geometries.
We note that $\ln 3/\ln 2\sim 1.58496$, $\ln 8/\ln 3\sim 1.89279$, 
$\ln 5/\ln 3\sim 1.46497$, 
and $\ln 4/\ln 3\sim 1.26186$. The question mark 
indicates the value that remains to be calculated, the Casimir energy
per unit length of a cylinder with the cross section of the shape of
a Koch snowflake. }
\label{tab-fdim}
\end{table}
\section{Discussion}

We list the fractal dimensions of the geometries discussed here
in Table~\ref{tab-fdim}. For the Sierpinski triangle and the 
Sierpinski carpet we have shown that the Berry conjecture holds true.
But, these are relatively trivial cases because the fractal dimensions
for these cases is equal to the respective topological dimensions.
For the inverse Sierpinski triangle, the inverse Sierpinski
carpet, and the inverse Vicsek fractal,
even though we encounter non-trivial fractal dimensions,
these dimensions (for area, perimeter, corner angle, and Casimir energy,)
turn out to be all equal. This, in turn, remarkably,
gives no room for the divergent terms 
in the Weyl expansion. Thus, in the absence of the divergent terms,
even though the Berry's conjecture holds in principle,
we can not conclude this to be a non-trivial verification of the conjecture.

Berry's conjecture was probably motivated for fractals like the 
Koch snowflake~\cite{Mandelbrot:1977ngf}, a simply connected domain,
in which the perimeter encloses a single continuously
connected region. The area and the corner angles of a 
Koch curve scales normally, but the perimeter scales like a 
fractal with a fractal dimension of $\ln 4/\ln 3\sim 1.26$. 
This difference in the scaling behavior between area and perimeter, we believe,
will be suitable for studying the Berry conjecture.
The methods presented here do not seem to yield the
Casimir energy of a Koch snowflake. 

The perimeter of a Koch snowflake scales like $P_k(a) = 4^nP_k(a/3^n)$,
which implies that for a Koch snowflake $\delta_1=\ln 4/\ln 3$.
In contrast, the area of a Koch snowflake $A_k(a)$ is given by
$A_k(a)=A_\Delta(a) +3A_k^\prime(a)$,
expressed in terms of a reduced area $A_k^\prime(a)$
that satisfies the recursion relation
$A_k^\prime(a) = A_\Delta(a/3) + 4A_k^\prime(a/3)$,
which leads to $A_k(a)= 8A_\Delta(a)/5$ and implies that for a Koch snowflake
$\delta_2=2$. Further, the corner angles $c_k(a)$
of a Koch snowflake satisfies $c_k(a) = 6\pi + 4c_k(a/3)$,
which leads to $c_k(a)=-2\pi$. Thus, for a Koch snowflake $\delta_0=0$.

It seems that it would be more appropriate to analyze 
the Berry conjecture for a Koch snowflake,
because of the fact that its perimeter scales like a fractal while its
area scales normally.
In the Sierpinski triangle, and in the inverse Sierpinski triangle,
the cavities are compartmentalized. These geometries are not simply 
connected. This feature in conjunction
with the idea of self-similarity was the key to our evaluation of
the Casimir energies of the fractal geometries considered here. 
Without the feature of compartmentalization we are unable to calculate
the Casimir energy of a Koch snowflake as yet.
Thus, we are unable to analyze the Berry conjecture for a Koch snowflake.

In the literature,
the Berry conjecture was formulated and has been studied 
for the distribution of the modes of a cavity. 
Casimir energy is directly related to the sum of all modes of
a cavity, and gets divergent contributions dictated by the distribution
of the modes for large frequencies. 
There seems to be an extensive literature on the Weyl law
and the associated Berry conjecture. We found the review in 
Ref.~\cite{Arendt:2009pm} very resourceful. Nevertheless,
to our knowledge, Berry's conjecture has not been addressed in the 
context of Casimir energies before.
The only exception seems to be the discussion in Ref.~\cite{Dunne:2012qe}
where fractal geometries are discussed in the context of heat kernels,
which is a powerful technique used to extract divergent terms in 
Casimir energy. Our method exploits the formalism of Green's functions,
that was described in the context of self-similar plates in
Ref.~\cite{Shajesh:2016yfq}, and its correspondence with the heat kernel
method is well known in the community.
In Ref.~\cite{Abalo:2010ah}, the method of mode summation was used,
so the connection with heat-kernel methods should not be so remote.
In our understanding, there is no direct overlap between the discussions we 
have presented here with those in Ref.~\cite{Dunne:2012qe}, but the 
connections should be worth pursuing. 

In the studies on distribution of modes, 
without concerning Casimir energy, it has been argued in 
Ref.~\cite{Brossard:1986fdh} that the dimensions of the regions 
and surfaces should be interpreted as the Minkowski-Bouligand dimension
instead of the Hausdorff-Besicovitch dimension as originally proposed by Berry.
This was further promoted in Ref.~\cite{Lapidus1988lp}.
Counterexamples involving domains that are not simply connected were 
presented in Ref.~\cite{Lapidus:1996dfc}, but it has been suggested that
the conjecture is expected to hold for simply connected fractals like the 
Koch snowflake. These conclusions seem to be in agreement with 
our results here in the context of Casimir energies.
No doubt, Berry's conjecture needs to be explored further.

\acknowledgments

We thank Kimball A. Milton for discussions, and for directing our attention to
the Casimir energy calculations using interior modes alone that 
was essential for this work. We thank Mathias Bostr\"{o}m for discussions.
We acknowledge support from the Research Council of Norway (Project No. 250346).
I.C.P. acknowledges support from Centro Universitario de la Defensa
(Grant No. CUD2015-12) and DGA-FSE (Grant No. 2015-E24/2).

\bibliography{biblio/b20170529-sierpinski-triangle}

\end{document}